\documentclass[prl,showpacs,twocolumn,amsmath,amssymb,superscriptaddress,floatfix]{revtex4-1}
\usepackage{color}
\usepackage{graphics}
\usepackage{epsfig}
\usepackage{amsmath}
\usepackage{amssymb}
\usepackage{amsthm}
\usepackage[mathscr]{eucal}

\newcommand{\bea}{\begin{eqnarray}}
\newcommand{\eea}{\end{eqnarray}}
\newcommand{\beas}{\begin{eqnarray*}}
\newcommand{\eeas}{\end{eqnarray*}}

\begin{document}

\title{Quarkyonic Chiral Spirals in the Nambu-Jona-Lasinio Approach}
\author{Bo Feng}
\affiliation{ School of Physics, Huazhong University of Science and Technology, Wuhan 430074, China}
\author{ Efrain J. Ferrer} 
\affiliation{ Department of Physics, University of Texas at El Paso, El Paso, TX 79968, USA}
 \author{Vivian de la Incera }
\affiliation{ Department of Physics, University of Texas at El Paso, El Paso, TX 79968, USA}

\begin{abstract}
We explore the  inhomogeneous QCD phases at finite density and temperature using a (3+1)-dimensional Nambu-Jona-Lasinio (NJL) model in the large $N_c$ limit with an additional attractive tensor-tensor interaction channel. For single modulated solutions, the problem reduces to solving the gap equation of a Chiral Gross-Neveu ($NJL_2$) theory, whose minimum solution is a periodic array of chirally twisted kinks. At zero temperature, the minimum solution reduces to the quarkyonic chiral spiral found in quarkyonic matter. The connection between the quarkyonic matter and our (3+1)-dimensional NJL model is rooted on the fact that the first reduces to (1+1)-dimensional QCD and the second to the $NJL_2$ theory, both of which have continuous chiral symmetry and asymptotic freedom. Our findings can be useful to obtain a reliable qualitative picture of the QCD phase diagram and the location of the QCD critical point in the technically challenging region of intermediate densities and temperatures.
\end{abstract}
\maketitle

\textbf{Introduction}. In this letter we provide a solution to a standing problem in the literature related to the existence of two different inhomogeneous solutions, each proposed as the most energetically favored ground state of large $N_c$ QCD in the region of intermediate densities and temperatures. One of these solutions, known as quarkyonic chiral spiral \cite{q-chiralspirals}, was found in quarkyonic matter \cite{quarkyonic-matter}.  The second solution was obtained using a (3+1)-dimensional (D) NJL theory with conventional scalar and pseudo-scalar interactions \cite{NickelPRD80}. The first solution is a complex order parameter that forms a chiral spiral of scalar and spin-one condensates. The second is a scalar condensate given by a real crystalline soliton solution. Both ground states are single-modulated and break translational symmetry in the direction of the modulation. In the present paper, by considering a (3+1)-D NJL model with an additional attractive channel that preserves the chiral symmetry of the theory, we demonstrate that the most favored ground state solution is a crystalline twisted chiral kink that reduces to the quarkyonic chiral spiral solution in the low temperature region. This modified NJL model contains the NJL theory of \cite{NickelPRD80}  as a particular case and is consistent with the quarkyonic matter results, suggesting that  the proposed NJL model can be more physically appropriate to describe the intermediate density region of QCD. 

Mapping all the phases of QCD is a goal intensely sought after by many theoretical and experimental efforts \cite{QCDreviews}. Thanks to the asymptotic freedom of the theory the most extremes regions of the QCD phase diagram in the temperature-density plane are perturbative and hence well understood;  they are the quark-gluon plasma in the high-temperature/low-density corner of the phase map and the Color-Flavor-Locked (CFL) superconducting phase on the opposite one.  At low temperatures and densities quarks are confined inside hadrons, whose interactions can be phenomenologically described by conventional nuclear physics.

However, somewhere in the region of intermediate temperatures and densities, one expects a phase transition to occur from a confined to a deconfined phase, where gluons and quarks liberate from hadrons. On the other hand, confined quarks have a large dynamical mass, while in the regions with small coupling the quark mass becomes small. Clearly, by increasing the temperature or the chemical potential, a chiral phase transition should occur from the phase of heavy quarks to that of light quarks. Nevertheless, it is not obvious a priori that the confinement and chiral phase transitions have to occur at the same temperature \cite{Aoki06}. 

Understanding the fundamental physics involved and the phases that realize in the strong-coupling region of the QCD phase map is nontrivial, as it requires using nonperturbative methods and effective theories. To begin with, one cannot use lattice QCD because of the sign problem at finite chemical potential, then the investigation of this region can only be done with the help of NJL-like \cite{NJL} or QCD-like effective models \cite{largeNQCD}.  To pin-down the correct physical description of this region, the theoretical results will have to be eventually contrasted with the outcomes of future experiments at various international facilities that plan to explore the intermediate-baryon-density and small-temperature region in the near future.

The strong-coupling region is even more technically challenging due to the possibility of inhomogeneous phases (IPs). Phases with inhomogeneous condensates (IC) have been found in the large N limit of QCD \cite{largeNQCD}, in quarkyonic matter \cite{q-chiralspirals}, and in NJL models \cite{NickelPRD80,largeN,PRD82-054009,PRD85-074002}. These phases seem unavoidable at intermediate densities and low temperatures. Indeed, while most NJL models predict a first-order chiral  transition with increasing density \cite{Klevansky92}, it is likely that the transition will occur via some intermediate state(s) characterized by inhomogeneous chiral condensates, because the density promotes an imbalance between quarks and antiquarks and an energy separation between them.  Since it costs energy hopping the antiparticles from the Dirac sea to pair with the particles at the Fermi sea, the system prefers instead to pair quarks and holes giving rise to ICs.

In this paper we are interested in exploring 1-D modulated ICs at finite temperature and density in a modified NJL model in the large Nc limit that takes into account new attractive channels. Previous studies of 1-D modulated ICs within (3+1)-D NJL models considered only the possibility of chiral and pion condensates \cite{NickelPRD80,largeN,PRD82-054009,PRD85-074002}. However, there are other attractive channels that respect the original chiral symmetry of the theory and can condense when some spatial symmetries are either spontaneously or explicitly broken. This is known to occur, for instance, in color superconductivity \cite{ferrer-incera} and quarkyonic matter \cite{ferrer-incera-sanchez} in the presence of a magnetic field. 

An important outcome from \cite{NickelPRD80} is the observation that the problem of finding
self-consistent IPs with single modulation can be reduced to finding the ground state solution in a (1+1)-D model. The relevance of this observation is the understanding that the pairing dynamics of single-modulated solutions is essentially (1+1)-D. Thanks to this effective dimensional reduction, one can use known results from (1+1)-D theories like the Gross-Neveu (GN) \cite{GN results,Thies-AnPhys} or the NJL$_2$ \cite{NJL2 results}, to tackle the problem of finding self-consistent phases with 1-D modulation in a (3+1)-D NJL model. 

One might naively think that since the chiral spiral solution of the (1+1)-D QCD theory \cite{q-chiralspirals} is directly related to the confinement of the interactions at the Fermi surface, one should not expect the same ground state in a (3+1)-D NJL model where confinement is absent. However, as will become clear later, this argument is not valid because the dimensional reduction of the pairing dynamics always connects the (3+1)-D NJL models to lower dimensional theories with asymptotic freedom. 

As we shall show below, once we extend the NJL model of Ref. \cite{NickelPRD80} to allow a tensor-tensor four-fermion interaction term that preserves the chiral symmetry of the theory, the quarkyonic chiral spiral solution becomes more energetically favored than the real, crystalline solution of \cite{NickelPRD80}. At the end of the paper,  we shall discuss, using symmetry arguments, what is the underlying physics making the connection between our modified NJL model and the confined quarkyonic matter.
 
\textbf{A modified (3+1)-D NJL Model}. We start with the following two-flavor chiral invariant NJL Lagrangian in (3+1) dimensions
\begin{eqnarray}\label{Lagrangian}
{\cal L}={\bar\psi}\left(\gamma^\mu(i\partial_\mu+\mu\delta_{\mu0}\right)\psi+G\left[({\bar\psi\psi})^2+({\bar\psi i\mathbb{\tau} \gamma_5 \psi})^2\right] 
\nonumber
\\
+G'\left[({\bar\psi \sigma^{\mu\nu}\psi})^2+ ({\bar\psi i\mathbb{\tau} \gamma_5 \sigma^{\mu\nu}\psi})^2\right] \qquad \qquad \qquad
\end{eqnarray}
 where $\mathbb{\tau}$ are Pauli matrices associated with the SU(2) flavor symmetry, $\sigma^{\mu\nu}=\frac{i}{2}[\gamma^\mu,\gamma^\nu]$, $\gamma^\mu$ is taken in the chiral representation, and $\psi$ is the 4$N_f N_c$ dimensional quark spinor with $N_f=2$. 
  
We consider the large $N_c$ limit, thereby avoiding the no-go theorem issue after the problem reduces to symmetry breaking in a lower dimensional theory \cite{Mermin-Wagner,Witten}. For simplicity, we are going to assume $G=G'$.  We are interested in 1-D modulated condensates and without lost of generality can take the modulation in the z-direction.  We shall consider the condensates 
$\langle  {\bar\psi}\psi\rangle=S(z)$ and $\langle{\bar\psi}\sigma^{03}\psi\rangle=D(z)$, because once the scalar condensate $S(z)$ selects a preferential direction, nothing prevent in principle the $D(z)$ condensate to exist. On the other hand, we can ignore the condensation of the second structures on each terms, because the pion condensate is known not to be favored \cite{NickelPRD80} and a similar result can be shown for the $\bar\psi i\tau\gamma_5 \sigma^{\mu\nu}\psi$ structure. 

In the mean-field  approximation the thermodynamical potential is 
\begin{eqnarray}
\Omega(T,\mu; M)
=-\frac{TN_cN_f}{V}\sum_n {\rm Tr}\ln\frac{1}{T}\left[i\omega_n+H_{MF}-\mu \right] \nonumber
\\
+\frac{1}{V}\int _{V}\frac{|M(z)|^2}{4G} \qquad \qquad \qquad \qquad
\end{eqnarray}
where 
\begin{eqnarray}
H_{MF}=-i\gamma^0\gamma^i\partial_i+\gamma^0\left[ d_{+} M(z)+d_{-}M^*(z) \right] \qquad \qquad
\label{Hamiltonian}
\end{eqnarray}
with $d_{\pm}= \frac{1 \pm \gamma^0\gamma^3}{2}$ and $-\frac{M(z)}{2G}=S(z)+i D(z)$. The trace acts on Dirac and coordinate space.
Using a proper regularization, one can express the thermodynamic potential in terms of the 
eigenvalues $E_n$  of $H_{MF}$ as 
\begin{eqnarray}
\Omega(T,\mu;M(z))=-\frac{TN_cN_f}{V}\sum_{E_n}\ln\left(2\cosh\left(\frac{E_n-\mu}{2T}\right)\right)\nonumber
\\
+\frac{1}{V}\int _{V}\frac{|M(z)|^2}{4G} + const. \qquad \qquad \qquad \qquad
\end{eqnarray}
The gap equation is obtained by minimizing the thermodynamic potential with respect to $M^*(z)$. It can be expressed in terms of the normalized eigenvectors $\psi_n(\mathbf{x})$ of $H_{MF}$  for the eigenvalues $E_n$ \cite{NickelPRD80},
\begin{equation}
M(z)=\frac{4GN_cN_f}{V}\sum_{E_n}\tanh\left(\frac{E_n-\mu}{2T}\right)\psi^\dagger_n(\mathbf{x})\frac{\partial{H_{MF}}}{\partial M^*(z)}\psi_n(\mathbf{x})
\label{gapeq}
\end{equation}
Following Ref. \cite{NickelPRD80}, we can take advantage of the translational invariance in the (x,y) plane,  which implies the conservation of the momentum component $P_{\bot}$ in that plane. The eigenfunctions of $H_{MF}$ can be labelled by the energy eigenvalues and the perpendicular momentum, $\psi_{\lambda, \mathbf{p}_{\bot}}$. With the help of a Lorentz boost they can be expressed in terms of the eigenfunctions $\psi_{\lambda,0}$ for $\mathbf{p}_{\bot}=0$. This allows to construct the eigenvalue spectrum from the subspace spanned by $\lbrace\psi_{\lambda,0}\rbrace$ (see \cite{NickelPRD80}  for details). In this subspace the Hamiltonian (\ref{Hamiltonian}) takes the form
\begin{equation}\label{nondiagonalH}
 H_{MF; 1D}=\left(
\begin{array}{cccc}
i\partial_z & 0 & M(z) & 0\\
0 & -i\partial_z & 0 & M(z)^*\\
M(z)^* & 0 & -i\partial_z & 0\\
0 & M(z) & 0 & i\partial_z
\end{array}\right)
\end{equation}
which can be block-diagonalized to
\begin{equation}\label{diagonalH}
H^\prime_{MF;1D}=\left(
\begin{array}{cc}
H_{1D}(M^*(z)) & 0\\
0 & H_{1D}(M^*(z))
\end{array}\right)
\end{equation}
where
\begin{equation}\label{1D-H}
H_{1D}(M^*(z))=\left(
\begin{array}{cc}
-i\partial_z & M^*(z)\\
M(z)& i\partial_z
\end{array}\right)
\end{equation}
In the superconductivity literature $H_{MF}$ is known as the Bogoliubov-de Gennes Hamiltonian \cite{superconductivity}. It is the Hamiltonian of the $NJL_2$ theory whose solutions were extensively studied in \cite{chiralspirals}. Then, our problem has been reduced to a direct product of two identical $NJL_2$ Hamiltonians, and hence we can identify the gap $\Delta$ in Ref. \cite{chiralspirals} with $M^*(z)$  and take advantage of the results found there. The gap equation of the theory (\ref{1D-H}) can be solved using a resolvent approach \cite{chiralspirals,PRD56-5050}. Its general solution is given by a periodic array of chirally twisted kinks \cite{chiralspirals}
\begin{eqnarray}\label{twistedkink}
M^*(z)=-\lambda e^{2iqz}A \frac{\sigma (\lambda Az+i \mathbf{K}' -i\theta/2 )}{\sigma (\lambda Az+i \mathbf{K}') \sigma (i\theta/2)} 
\nonumber
\\
\times \exp[i\lambda Az(-i \zeta(i \theta/2)+i \mathrm{ns}(i \theta/2))+i \theta\eta_{3}/2]
\end{eqnarray}
where $A=A(\theta,\nu)=-2i \mathrm{sc}(i\theta/4;\nu) \mathrm{nd}(i\theta/4; \nu)$;  $\mathrm{sc}=\mathrm{sn}/\mathrm{cn}$, $nd=1/dn$ are Jacobi elliptic functions, and $\sigma$ and $\zeta$ are the Weierstrass sigma and zeta functions, chosen to have real and imaginary half periods: $\omega_1=\mathbf{K}(\nu)$, and $\omega_3=i\mathbf{K}'(\nu)\equiv i\mathbf{K}(1-\nu)$, with $\mathbf{K}(\nu)$ the complete elliptic integral. Both periods are controlled by a single (real) elliptic parameter $0 \leq \nu \leq 1$. The parameter $\lambda$ sets the overall scale of the condensate, and $1/\lambda$ sets the length scale of the crystal.  Hence, the general solution is specified by four parameters: a scale $\lambda$, a phase parameter $q$, an angular parameter $\theta$ ($\theta \in [0,4\mathbf{K}']$), and the elliptic parameter $\nu$. 

For the extreme values of $\theta$, the twisted kink solution reduces to a chiral spiral
\begin{equation}\label{spiralsol}
M=\lambda e^{-2iqz}
\end{equation}
At zero temperature the energetically favored solution is the spiral with $q=\mu$ \cite{chiralspirals}. This is the same quarkyonic chiral spiral solution found in quarkyonic matter at zero $T$ \cite{q-chiralspirals}. This result indicates that the (3+1)-D NJL model \eqref{Lagrangian} encompasses the same physics as the one taking place in a patch of the Fermi surface in the quarkyonic matter. This is encouraging, as one may then use this NJL theory to explore other regions of the parameter space that would be rather difficult to describe with QCD effective models with confined gluon propagators. Here again one can take advantage of the connection with the $NJL_2$  model and the existing results for this lower dimensional theory \cite{chiralspirals}.

\textbf{Ginzburg-Landau Analysis.} We are particularly interested in exploring the region near the chiral transition curve to determine the form of the most energetically favored solution there. With this aim, we can use the Ginzburg-Landau (GL) expansion of the thermodynamical potential in powers of the condensate and its derivatives,
\begin{align}
&\Psi_{GL}(M^*)-\Psi_{GL}(0)& \notag \\
=&2\big\{\alpha_2 \vert M^* \vert^2+ \alpha_3 \mathrm{Im}(M^*M') + \alpha_4 (\vert M^* \vert^4 \notag \\
&+\vert M^{* \prime} \vert^2)+\alpha_{5} \mathrm{Im}((M^ {\prime \prime*}-3 \vert M^* \vert ^2 M^*)M') \notag \\
&+ \alpha_6 [2 \vert M^* \vert^6+8\vert M^* \vert^2\vert M^{\prime *} \vert^2 +2\mathrm{Re} ((M^{\prime *})^2 M^2) \notag  \\
&+ \vert {M}^{\prime\prime *} \vert^2] +...\} \label{GLexpansion}
\end{align} 
where the coefficients $\alpha_n (T, \mu)$ are functions of $T$ and $\mu$ given in \cite{chiralspirals}. Their explicit expressions are not relevant for our analysis. Expression \eqref{GLexpansion} is just twice the $NJL_2$ GL functional found in \cite{chiralspirals} because the (3+1)-D Hamiltonian  \eqref{diagonalH} is the direct product of two $H_{1D}(M^*(z))$, so its GL functional is the sum of two $NJL_2$ GL functional.

At each given order of the GL expansion, one can minimize the GL functional with respect to $M(z)$ to find the GL equation for the condensate at that order. Expanding up to $\alpha_3$ we obtain
\begin{equation}
M^{\prime *}- i \frac{2\alpha_2}{\alpha_3}M^*=0 \Rightarrow M^* = \lambda e^{i \frac{2\alpha_2}{\alpha_3}z} 
\end{equation}
which is again a chiral spiral solution. One can verify that the mimimun solution at each order of the expansion is always a chiral spiral \cite{chiralspirals}. Therefore, the chiral spiral is energetically preferred over the real crystalline solution of \cite{NickelPRD80} in the region close to the chiral transition. 

Notice that the GL functional \eqref{GLexpansion}, being twice the functional of  $NJL_2$, corresponds to a lower dimensional theory that has asymptotic freedom and continuous chiral symmetry \cite{chiralspirals}. These are the same symmetries of the (1+1)-D QCD theory that describes the confined interactions in the Fermi surface of the quarkyonic matter.

\textbf{Discussion.} In this letter we have unveiled a connection between quarkyonic matter and a modified (3+1)-D NJL model that has an attractive chiral-invariant tensor-tensor four-fermion interaction channel in addition to the conventional scalar and pseudo scalar terms. Condensation in the extra channel is favored once the spatial symmetry is spontaneously broken by the scalar condensate. For single-modulated condensates, the pairing dynamics of our (3+1)-D NJL model becomes (1+1)-D and is described by the $NJL_2$  theory. In the case of quarkyonic matter, the relevant pairing dynamics is described by (1+1)-D QCD.   Since the symmetries of these two lower dimensional theories are the same, it is logical to expect that they posses the same ground state. As a matter of fact, the ground state of our (3+1)-D NJL model is a crystalline chirally twisted kink, which becomes equal, at zero temperature, to the chiral spiral solution found in the patches of the Fermi surface in quarkyonic matter. 

 It is now easy to understand why the most favored solution  for the NJL model \eqref{Lagrangian} differs from the one found in \cite{NickelPRD80}. The reason is that these two NJL models are connected to different lower dimensional theories which have different symmetries. The (3+1)-D Hamiltonian of the present case is connected to the Chiral GN or $NJL_2$ theory. In contrast, the Hamiltonian of Ref. \cite{NickelPRD80}  is connected to the GN model. The chiral symmetry is continuous in the first, but it is discrete in the second. For the first model the Hamiltonian reduces to the direct product of two $NJL_2$ Hamiltonians, so the GL expansion is twice the $NJL_2$ GL expansion. In the second case, the Hamiltonian reduces to the direct product of $H_{1D}(M(z))$
and $H_{1D}(M^*(z))$, so the terms with odd-labelled coefficients in the GL expansion cancel out. Without these terms the minimum solution is never a chiral spiral, but a real function as shown in \cite{NickelPRD80}. 

The connection of the proposed NJL model with the quarkyonic matter is rooted on the dimensional reduction of the dynamics of fermions in the Fermi surface.  The fermions close to the Fermi surface are the relevant degrees of freedom to produce the particle-hole condensate that yields the inhomogeneous phase. Thus, despite the apparent differences between the initial (3+1)-D quarkyonic matter theory and the proposed (3+1)-D NJL model, one with confinement and the other without it, each of them reduces to a lower dimensional theory with confinement and the same continuous chiral symmetry, thereby producing the same ground state.


The proposed NJL model can be useful to explore the region in the $\mu-T$ plane occupied by the inhomogeneous phase, the kind of transitions that limit this region, and the possible existence and location of the QCD critical point.  

{\bf Acknowledgments:} The work of EJF and VI has been supported in part by DOE Nuclear Theory grant DE-SC0002179.

\end{document}